\documentclass[10pt]{iopart}
\usepackage{microtype}
\usepackage{iopams}
\usepackage{graphicx}
\usepackage{cite}
\usepackage{color}

\newcommand{\be}{\begin{equation}}
\newcommand{\ee}{\end{equation}}
\newcommand{\ra}{\rightarrow}

\newcommand{\scatt}{{\rm scatt}}
\newcommand{\prob}{P}
\newcommand{\tot}{{\rm tot}}
\newcommand{\stat}{{\rm stat}}
\newcommand{\erfc}{{\rm erfc}}
\newcommand{\mE}{\mathcal{E}}

\begin{document}

\title[Fluctuation relations without large deviations]{Fluctuation relations without uniform large deviations}

\author{Giacomo Gradenigo,$^1$ Alessandro Sarracino,$^1$ Andrea Puglisi,$^1$ and Hugo Touchette$^{2,3}$}

\address{
$^1$\ CNR - ISC and Dipartimento di Fisica, Universit\`a di Roma ``La Sapienza'', p.le A. Moro 2, 00185, Roma, Italy\\
$^2$\ National Institute for Theoretical Physics (NITheP), Stellenbosch 7600, South Africa\\
$^3$\ Institute of Theoretical Physics, University of Stellenbosch, Stellenbosch 7600, South Africa
}

\eads{
\mailto{ggradenigo@gmail.com},
\mailto{alessandro.sarracino@roma1.infn.it},
\mailto{andrea.puglisi@roma1.infn.it},
\mailto{htouchet@alum.mit.edu}
}

\begin{abstract}
We study the fluctuations of a stochastic Maxwell-Lorentz particle
model driven by an external field to determine the extent to which
fluctuation relations are related to large deviations. Focusing on the
total entropy production of this model in its steady state, we show that, although the
probability density of this quantity globally satisfies (by
definition) a fluctuation relation, its negative tail decays
exponentially with time, whereas its positive tail decays slower than
exponentially with time because of long collision-free
trajectories. This provides an example of physical system for which
the fluctuation relation does not derive, as commonly thought, from a
probability density decaying everywhere exponentially with time or, in
other words, from a probability density having a uniform large
deviation form.
\end{abstract}


\date{\today}

\section{Introduction}

The fluctuation relation (FR) is an important result of nonequilibrium
statistical mechanics, expressing a general asymmetry for fluctuations
of systems driven in nonequilibrium steady states. Such an asymmetry
typically applies for observables $A_t$ integrated over a time $t$ and
implies that 
\be \frac{P(A_t/t=a)}{P(A_t/t=-a)}=e^{cta+o(t)},
\label{FR}
\ee 
where $P(\cdot)$ denotes the probability density function (pdf),
$c$ is a constant that does not depend on $t$ nor $a$, and $o(t)$
stands for sublinear, i.e., sub-extensive corrections in $t$. The
exponential dominance of positive fluctuations over negative ones
expressed by this result has been widely studied for chaotic and
stochastic systems
\cite{ES94,gallavotti1995,gallavotti1995a,kurchan1998,lebowitz1999},
as well as for different physical observables of these systems, such
as the entropy production, particle currents, and work-like
quantities~\cite{visco2005,villavicencio2012,zon2003a,zon2004a,imparato2006a,sarracino2010,gradenigo2012b};
see \cite{evans2002,harris2007,marconi2008} for reviews. Observables
that satisfy the FR of Eq.~(\ref{FR}) have also been measured
experimentally, e.g., in manipulated Brownian particle experiments and
noisy electrical circuits
\cite{ciliberto2004,garnier2005,douarche2006,andrieux2007a,joubaud2007}.

For all of these systems, the exponential form of the FR is known to arise because of two fundamental properties of the pdf $P(A_t/t)$: (i) it satisfies a \emph{large deviation principle} (LDP) \cite{dembo1998,ellis1985,touchette2009}, i.e., 
\be
\lim_{t\ra\infty}-\frac{1}{t}\ln P(A_t/t=a)=I(a)
\label{eqldp1}
\ee
or equivalently,
\be
P(A_t/t=a)= e^{-tI(a)+o(t)},
\label{eqldpa}
\ee
and (ii) the so-called \emph{rate function} $I(a)$ has the symmetry property:
\be
I(-a)-I(a)=ca.
\label{eqsym1}
\ee
These two conditions, with $I(a)$ different from $0$ and $\infty$, are known to be sufficient for $A_t/t$ to satisfy an FR (see, e.g., \cite{touchette2009} and references therein).

Our goal in this paper is to show that the FR can
also arise from a pdf that does not have a leading
exponential scaling in $t$, as in Eq.~(\ref{eqldpa}), and so does not
satisfy a ``standard'' or ``classical'' LDP. By
considering a stochastic Maxwell-Lorentz gas driven by an external
field, we show that the entropy production calculated over a time $t$ in the steady state
satisfies an FR, even though the far positive tail of
its pdf scales exponentially with $\sqrt{t}$ rather than $t$.  This
provides a physical example for which an FR arises not from a single,
\textit{uniform} LDP as above, but from different large deviation
scales (here two scales), which can be fully characterized only by
explicitly calculating the $o(t)$ correction term in
Eq.~(\ref{eqldpa}). 

In this model, the $t$ time scale is physically related to ``normal'' trajectories of the gas' particles involving many collisions, whereas the $\sqrt{t}$ time scale is related to long
ballistic trajectories lasting for a time proportional to $t$, 
which lead under the external field to large positive fluctuations of the entropy
production scaling as $t^2$. The former type of trajectories or regime is discussed in Section 
\ref{sec:LDPentropy}, while the latter is discussed in Section \ref{breakdown}.
The relevance of our results for more general collisional
models is discussed in the concluding section of the paper.

It is important to note that the features of the model that we study
are different from those studied in the context of so-called
\emph{extended} FRs \cite{zon2003} and of boundary-term effects in FRs
\cite{puglisi2006a,harris2006,bonetto2006,zamponi2007}. In those
studies, the LDP condition of Eq.~(\ref{eqldpa}) is satisfied, but the
associated rate function does not satisfy the symmetry
(\ref{eqsym1}). Our results are also not related to \emph{anomalous}
FRs, which arise when there is no LDP because the pdf
of interest has power-law tails
\cite{touchette2007,touchette2009b,chechkin2009}. 

The model that we study has an LDP, which interestingly preserves the exact exponential scaling form
of the FR without it being uniformily exponential. A similar behaviour was
found recently in a non-Markovian random walk \cite{harris2009}.
From a more general perspective, non-standard LDPs are also found in systems showing phase
transitions (e.g., the Ising model at its critical point \cite{touchette2009}), noisy 
dynamical systems with non-isolated attractors \cite{bouchet2012},
quantum quenches~\cite{gambassi2012}, as well as disordered systems \cite{hollander2000}.

\section{Model and relevant stochastic variables}
\label{sec:model}

The model that we consider is a stochastic Maxwell-Lorentz gas previously analyzed in
\cite{alastuey2010,gradenigo2012a}, consisting of a probe particle of mass $m$ whose
velocity $v$ changes because of collisions with
particles from a bath and acceleration due to an external force field. 

The collision
process is represented by interactions with a bath of scatterers of mass
$M$ equilibrated at temperature $T$. Collisions with the scatterers 
change instantaneously the probe's velocity from $v$ to $v'$ according to the rule
\begin{equation}
v'=\gamma v + (1-\gamma) V, \qquad \gamma=\frac{\zeta-\alpha}{1+\zeta},
\end{equation}
where $\alpha\in[0,1]$ is the restitution coefficient ($\alpha=1$ for
elastic collisions), $\zeta=m/M$ is the mass ratio, and
$V$ is the velocity of the scatterer, taken to be distributed according to
the Gaussian pdf:
\begin{equation}
P_{\scatt}(V)= \sqrt{\frac{q}{\pi}}\, e^{- q V^2}
\label{eq:gaussian}
\end{equation}
with $q=M/(2T)$.  The fact that the pdf above is independent of the probe's velocity
means physically that the scatterers do not keep any memory of their collisions
 with the probe and, so, that the bath
of scatterers thermalizes rapidly, relative to the typical interaction
time, with the probe particle.

During a time $\tau$ between two consecutive collisions, the probe performs a deterministic
acceleration under the influence of an external field $\mE$. 
To simplify the model, we assume
that $\tau$ is taken for each flight from the exponential
pdf,
\begin{equation}
P_\tau(\tau)= \frac{1}{\tau_c} e^{-\tau/\tau_c},
\label{eqexp1}
\end{equation} 
which is independent of the relative velocity of the particles. This is a simplification
compared to collisional models with hard-core interactions, where the rate of collisions is
proportional to the relative velocity $|v-V|$; see, e.g., \cite{ernst,baldassa}.

Under the collision rule and acceleration defined above, the evolution of the velocity pdf $P(v,t)$ of the probe particle is described by the following linear Boltzmann equation:
\begin{eqnarray}
\label{prob}
\hspace{-2.0cm}
\tau_c \partial_t  P(v,t)+ \tau_c  \mE \partial_v P(v,t) = -P(v,t)+ \frac{1}{1-\gamma}\int\! du\,
P(u,t) P_{\scatt}\left(\frac{ v-\gamma u} {1-\gamma}\right), 
\end{eqnarray}
where $\tau_c$ appears as the mean collision time. Various properties
of this integro-differential equation are discussed in
\cite{alastuey2010,gradenigo2012a}. In particular, for the case $M>m$
and $\zeta = \alpha$, i.e., $\gamma=0$, which implies the simple
collision rule $v'=V$, the stationary Boltzmann
equation can be solved analytically to find for the steady state
\begin{eqnarray}
\hspace{-2cm} 
  P_\stat(v) = b \sqrt{\frac{q}{\pi}} \int_0^\infty du~e^{ - q(u-v)^2 - b u }=\frac{1}{2} b\, e^{\frac{b (b-4 q v)}{4 q}} {\erfc}\left(\frac{b-2 q v}{2
   \sqrt{q}}\right),
  \label{eq:stationary}
\end{eqnarray} 
with $b=1/(\mE \tau_c)$. This is the case that we consider throughout
this paper. For the general case $\gamma \neq 0$, an explicit form of
the solution is not available, although it can be written as a series
expansion in Fourier space \cite{gradenigo2012a}.

It is worth noting that the collisional model described by Eq.~(\ref{prob}) belongs to a more
general class of models recently discussed in~\cite{barbier2012},
whose collision integral includes a
term of the form $|v-u|^\nu$ in the kernel and a
scatterers' pdf generalised to $P_{\scatt}(V) \sim
e^{-|V|^\mu}$ (here we consider $\nu=0$ and $\mu=2$).  In~\cite{barbier2012} it is shown
that the parameter space $(\nu,\mu)$ is divided by a ``transition
line'' in two regions: $\mu<\nu+1$ corresponds to models for which the
interaction with the scatterers is strong enough to guarantee thermalization, while for $\mu>\nu+1$, which is the case
considered here, the accelerated
particle takes a stationary pdf which is far from that of the
scatterers. The model that we study is also in the
class of the original Lorentz gas~\cite{lorentz} and the frequently-studied Sinai billiard \cite{sinai}, for which $\nu=1$ (hard spheres) and
$\mu=\infty$ (scatterers at rest).

\section{Total entropy production}
\label{sec:entropydec}

The quantity that we study for the purpose of comparing the FR and the
LDP is the total entropy production $\Delta s_{\tot}$ associated with the velocity $v(t)$. This quantity is defined in the standard way as

\begin{equation} 
\Delta s_{\tot}(t) = \ln \frac{\prob(\{v(s)\}_0^t)}{\prob(\overline{\{v(s)\}_0^t})},
\label{epdef}
\end{equation}
where $\prob(\{v(s)\}_0^t)$ and
$\prob(\overline{\{v(s)\}_0^t})$ are, respectively, the stationary pdf of a
path $\{v(s)\}_0^t$ spanning the time interval $[0,t]$ and of the
time-reversed path $\overline{\{v(s)\}_0^t}=\{-v(t-s)\}_0^t$~\cite{lebowitz1999}. From this definition, it is easy to check (see \cite{lebowitz1999,marconi2008}) that the stationary probability of observing $\Delta s_{\tot}=a$ obeys
\begin{equation}
P(\Delta s_{\tot}=a)= e^a\, P(\Delta s_{\tot}=-a),
\end{equation}
so that
\be
\frac{P(\Delta s_{\tot}/t=s)}{P(\Delta s_{\tot}/t=-s)}=e^{ts}
\label{eqpr1}
\ee
for all $t>0$ and $s\in\mathbb{R}$. Therefore, $\Delta s_{\tot}/t$ satisfies as announced an FR with $c=1$, which holds \emph{exactly} in this case as there is no $o(t)$-corrections to Eq.~(\ref{FR}).

At variance with this result, we show in the next section that $P(\Delta s_{\tot}/t=s)$ itself is not everywhere exponential with $t$, which means that the FR in this case does not derive, following the introduction, from an LDP having a uniform scale or \emph{speed} $t$, but from a non-trivial combination of LDPs showing different scalings with time.

To derive these LDPs, we use in the next section a result of \cite{gradenigo2012a} showing that $\Delta s_{\tot}$ can be decomposed as
\begin{equation}
\Delta s_{\tot}(t) = \frac{W}{\theta} +B
\label{eq:entropy2}
\end{equation} 
where
\begin{equation}
W(t) = m\,\mE \int_{t_1}^{t_n} v(s) ds = m\,\mE  \sum_{i=1}^{n-1} x_i
\label{eq:work}
\end{equation} 
is the work done by the external field between the first and the last
collisions, and
\begin{equation}
B=\frac{m}{2
  \theta}[v^2(t_1^-)-v^2(t_n^+)] + \ln
  \frac{P(v(0))}{P(-v(t))}
\end{equation} 
is a boundary term involving only velocities close to the start and to
the end of the time interval $[0,t]$.  In the above formulas, $n$ is
the (random) number of collisions up to time $t$, $\theta=T \zeta$ is
the energy scale corresponding to equipartition with the scatterers
(which is not reached because of the external field and the
dissipation in collisions), while $v(t_i^+)$ and $v(t_i^-)$ are the
velocities of the particle after and before the $i$-th collision,
respectively. Moreover,
\begin{equation}
x_i =\int_{t_i}^{t_{i+1}} v(s) ds
\end{equation}
represents the probe's displacement between the times $t_i$ and
$t_{i+1}$.  For $\gamma=0$, the post-collisional
velocities $v(t_i^+)$ are extracted from the Gaussian
pdf (\ref{eq:gaussian}), whereas the pre-collisional
velocities are distributed according to the pdf (\ref{eq:stationary}).

\section{Center LDP for the total entropy production}
\label{sec:LDPentropy}

Rate functions of random variables are often obtained by calculating their corresponding scaled cumulant
generating function (SCGF) \cite{touchette2009}. For the total entropy production, the SCGF is defined by the limit
\begin{equation}
\lambda_{\Delta s_{\tot}}(k)=\lim_{t\rightarrow\infty}
\frac{1}{t}\ln\langle e^{k \Delta s_{\tot}}\rangle.
\label{lds}
\end{equation}
Following the G\"artner-Ellis Theorem \cite{dembo1998,ellis1985,touchette2009}, it is possible to obtain the rate function $I(s)$ of $\Delta s_{\tot}/t$, defined by
\begin{equation}
I(s)=\lim_{t\ra\infty} -\frac{1}{t}\ln P(\Delta s_{\tot}/t=s),
\label{eqldp3}
\end{equation}
provided $\lambda_{\Delta s_{\tot}}(k)$ is differentiable and steep in the interior of its domain of existence.\footnote{The steepness condition means essentially that the derivative of the SCGF diverges at boundary points of its domain; see Examples 3.3 and 4.8 of \cite{touchette2009} for a complete explanation.} Under these conditions, we then have that $I(s)$ is the Legendre-Fenchel (LF) transform of $\lambda_{\Delta s_{\tot}}(k)$, i.e., 
\begin{equation}
I(s) = \max_{k\in\mathbb{R}}\{sk-\lambda_{\Delta s_{\tot}}(k)\}.
\label{legendre}
\end{equation}

The calculation of $\lambda_{\Delta s_{\tot}}(k)$ is simplified here
thanks to the decomposition of Eq.~(\ref{eq:entropy2}) and the fact that the work $W(t)$,
which is an integral of a random process over the time interval $[0,t]$, is asymptotically independent of the boundary term $B$, which involves only the limits of this integral. Defining the SCGF of the work by
\be
\lambda_W(k)=\lim_{t\rightarrow\infty} \frac{1}{t}\ln\langle e^{k W/\theta}\rangle
\ee
and that of the boundary term by
\be
\lambda_{B}(k)=\lim_{t\rightarrow\infty} \frac{1}{t} \ln\langle e^{k B }\rangle, 
\ee
we must then have
\be
\lambda_{\Delta s_{\tot}}(k)=\lambda_W(k)+\lambda_B(k).
\label{eq:DStot-expr}
\ee

The calculation of $\lambda_W(k)$ is carried out in \ref{appscgf} with the result,
\be
\lambda_W(k) =  \frac{G_x(k)-1}{\tau_c},
\label{eq:SMnew}
\ee
where $G_x(k)=\langle e^{\frac{k m \mE}{\theta} x} \rangle$ is the generating function of the displacements $x_i$ appearing in the expression of the work, Eq.~(\ref{eq:work}). This generating function cannot be obtained in closed form; however, it can be evaluated numerically, and from its integral representation, shown in Eq.~(\ref{eq:deltax}), we find that $\lambda_W(k)<\infty$ for $k\in(-1,0]$ and $\lambda_W(k)=\infty$ otherwise. The domain of existence of $\lambda_W(k)$ is thus $(-1,0]$. Moreover, from Eq.~(\ref{eq:deltax}) we obtain the FR-like symmetry
\be
\lambda_W(k)=\lambda_W(-k-1)
\label{eqsym2}
\ee 
for all $k\in\mathbb{R}$ if we accept the equality $\infty=\infty$.

The SCGF of the boundary term $B$ is calculated in a different way by rewriting it as the sum $B=b_1+b_n$ of two asymptotically independent
terms:
\begin{eqnarray}
b_1 = \frac{m}{2\theta} v^2(t_1^-) + \ln P(v(0)) \nonumber\\
b_n = - \frac{m}{2\theta} v^2(t_n^+) - \ln P(-v(t)). 
\label{bn}
\end{eqnarray}  
Since these terms do not scale extensively with $t$, their SCGFs must vanish when it exists, so that we only need to determine their domain of existence. This is done in~\ref{appscgf2}. The result that we obtain is that the SCGF of $b_1$ converges for $k\in (-\infty,0]$, while that of $b_2$ converges for $k\in (-1,0]$. Combining these domains, i.e., taking their intersection, we then find 
\begin{equation}
\lambda_B(k)=
\left\{
\begin{array}{lll}
0 & & k\in (-1,0] \\
\infty & & \textrm{otherwise}.
\end{array}
\right.
\end{equation}
Thus $\lambda_B(k)$ has the same domain as $\lambda_W(k)$, which means
that we can finally write \be \lambda_{\Delta s_{\tot}}(k)=\lambda_W(k) \ee with $\lambda_W(k)$ given again by
Eqs.~(\ref{eq:SMnew}) and (\ref{eq:deltax}). This shows that only the
work plays a role in the SCGF of the entropy production --
the boundary term is irrelevant.

\begin{figure}[t!]
\begin{center}
\includegraphics[width=0.9\columnwidth,clip=true]{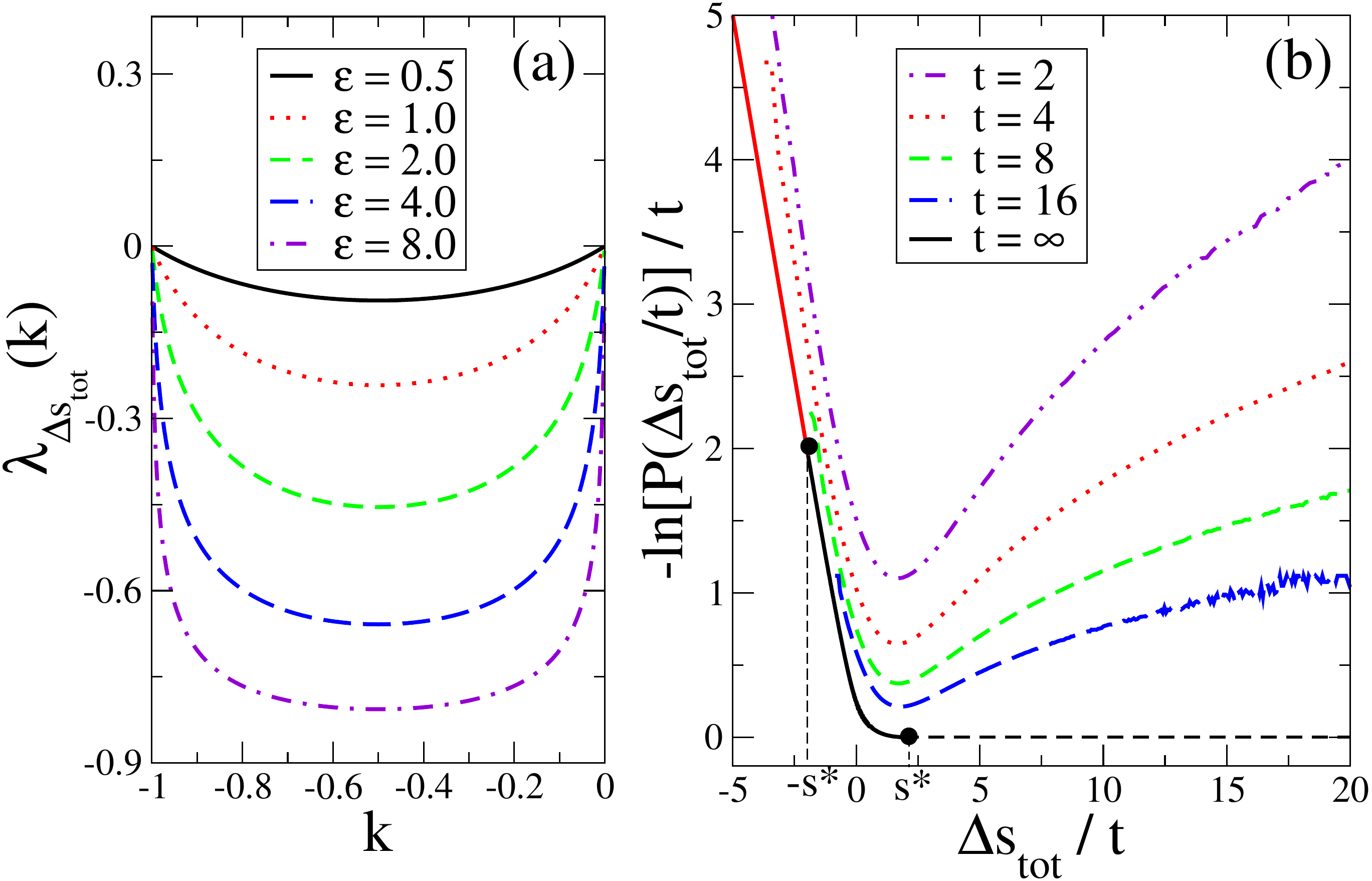}%
\caption{(Color online) (a) SCGF for the total entropy production,
  Eq.~(\ref{eq:DStot-expr}), for the parameters $m=1, M=2, \tau_c=1,
  T=1, \gamma=0$ and various values of the field $\mE$. (b)
  Numerical estimates of the rate function $I(s)$ obtained at
  different times measured in units of the mean collision time
  $\tau_c$.  The numerical simulations were obtained, following the method described in \cite{gradenigo2012a}, with the
  parameters $m=1, M=2, T=1, \tau_c=1, \gamma=0, \mE=1$. The
  $t=\infty$ black line represents the rate function obtained from the Legendre transform of the SCGF; 
the red line shows the part $I(s)=-s$ for $s\le -s^*$, while the dotted line shows the part $I(s)=0$ for $s\geq s^*$.} 
\label{figure1}
\end{center}
\end{figure}

In Fig.~\ref{figure1}(a) we plot $\lambda_{\Delta s_{\tot}}(k)$ by numerically computing the
integral in Eq.~(\ref{eq:deltax}) for several values of the
field $\mE$. Notice that, as expected from the symmetry (\ref{eqsym2}), $\lambda_{\Delta s_{\tot}}(k)$ is symmetric with respect to $k=-0.5$, and that its right-derivative at $k=-1$ and left-derivative at $k=0$ increase in magnitude as the field $\mE$ is increased. From Eq.~(\ref{eq:deltax}), we actually find
\be
\lambda'_{\Delta s_{\tot}}(k=0^-)=-\lambda'_{\Delta s_{\tot}}(k=-1^+)=\frac{m\tau_c\mE^2}{\theta}.
\label{eqslope1}
\ee
This value is important for our discussion and will be denoted next by $s^*$.

Having the SCGF of the entropy production, we can now in principle obtain its LDP using the G\"artner-Ellis Theorem. However, the problem arises that the SCGF is not steep, so it is not clear whether the rate function $I(s)$ is the LF transform of the SCGF. What is known is that this transform yields the correct $I(s)$ for $s$ in the image of the derivative of $\lambda_{\Delta s_{\tot}}(k)$, which here corresponds to $(-s^*,s^*)$. Over this interval, we can indeed express the rate function as the Legendre transform of the SCGF,\footnote{See Example 3.3 and Secs.~4.1 and 4.4 of \cite{touchette2009} for more details.} so that
\be
I(s)=k\lambda'_{\Delta s_{\tot}}(k)-\lambda_{\Delta s_{\tot}}(k)
\ee
for 
\be
s=\lambda'_{\Delta s_{\tot}}(k)\in (-s^*,s^*).
\ee
The result of this Legendre transform is shown in Fig.~\ref{figure1}(b). It is worth noticing that $I(s^{*})=0$. Moreover, from the symmetry of Eq.~(\ref{eqsym2}), we readily obtain
\be
I(-s)-I(s)=s,
\ee
so that the original FR symmetry of Eq.~(\ref{eqsym1}) is satisfied for $s\in (-s^*,s^*)$. Over this interval, we have in fact
\be
P(\Delta s_{\tot}/t=s)=e^{-t I(s)+o(t)},
\ee
so that both the LDP \emph{and} the FR for the entropy production are exponential in $t$, in agreement with the scenario outlined in the introduction.

Outside the centered interval $(-s^*,s^*)$, one cannot be
certain that the rate function is the LF transform of the SCGF because
this transform is one-to-many for non-steep functions \cite{touchette2009}. Other arguments
are given in the next section to obtain the rate function outside this
interval, with the result that $P(\Delta s_\tot/t=s)$ should decay with $\sqrt{t}$ rather than $t$
for $s>s^*$, implying that $I(s)=0$ for $s\geq s^*$ and $I(s)=-s$
for $s\leq -s^*$. This is consistent with the LF transform of
$\lambda_{\Delta s_{\tot}}(k)$ and is supported by simulation
results presented in Fig.~\ref{figure1}(b). 

\section{Far tail LDPs}
\label{breakdown}

The typical trajectories of the system that give rise to fluctuations of $\Delta s_\tot$ close to its typical value $s^*$ are those that involve many collisions (of the order of $t/\tau_c$) over the time $t$. To study the large fluctuations of this quantity away from $s^*$ -- at least positive ones -- we must therefore study long ballistic trajectories of the probe particle involving few collisions. Physically, these trajectories, exponentially distributed in time, have the effect of producing a
``slower-than-exponential'' or ``fat'' LDP in the far positive tail of the
entropy production pdf.\footnote{Note that this fat tail
  does not induce any anomalous diffusion, i.e., one has
  $\langle [x(t)-\langle x(t) \rangle]^2\rangle \sim t$ at large
  times~\cite{gradenigo2012a,GSVV12}.}

To see this, consider the extreme case where no collision occurs in $[0,t]$ so that $\tau=t$. Then, as shown in \cite{gradenigo2012a}, the entropy production is exactly given by
\be
\Delta s_\tot = \ln \frac{P(v(0))}{P(-v(t))},
\ee
where $v(0)$ is distributed according to the stationary pdf of Eq.~(\ref{eq:stationary}) and $v(t)=v(0)+\mE t$. From the expression of the stationary pdf, it is easy to see that its negative tail has the form of a Gaussian, which implies
\be
P(-v(t))=P_{\stat}(-v(0)-\mE t)\sim e^{-q(v(0)+\mE t)^2},
\ee
for $v(t)\gg 1$ and $\mE>0$, so that
\be
\Delta s_{\tot} \sim q \mE^2 t^2 =\frac{m}{2\theta} \mE^2 t^2
\label{eqasymps1}
\ee
if we retain in the entropy production only the dominant positive term that scales with $t$. The probability of this ``no-collision'' event is given by the exponential pdf of $\tau$ shown in 
Eq.~(\ref{eqexp1}), so that by change of variables we find 
\be
P(\Delta s_{\tot}/t=s)\sim e^{-\kappa \sqrt{t}\sqrt{s}},\qquad
\kappa =\sqrt{\frac{2\theta}{m\mE^2\tau_c^2}}
\label{eqsubldp1}
\ee
with subexponential corrections in $\sqrt{t}$.

The same result is obtained if we consider a long ballistic trajectory with one collision at either end of $[0,t]$ or two collisions close to the start and end of this time interval. In the latter case, the dominant part of the entropy production is the work done over the displacement
\be
x= v t +t^2\mE/2\sim t^2\mE/2,
\ee
which leads us to the same asymptotic for $\Delta s_\tot$ as that shown in (\ref{eqasymps1}) and, consequently, the same LDP shown in (\ref{eqsubldp1}). Hence the extreme positive fluctuations of the entropy production are governed by a ``fat'' LDP, which scales with $\sqrt{t}$ rather than $t$. This explains why the SCGF of the entropy production diverges for $k>0$: the stretched exponential tail of the pdf~(\ref{eqsubldp1}) is subleading compared to the exponential term in the SCGF (\ref{lds}), and so cannot compensate for the divergence of this term for $k>0$.

These simple arguments cannot be used to determine the precise value $s$ at which the change of scaling from $t$ to $\sqrt{t}$ occurs. However, if we assume that there is no additional scaling and that the pdf of the entropy production is unimodal, then the divergence of $\lambda_{\Delta s_{\tot}}(k)$ for $k>0$ together with the value $s^*$ of its left-derivative at $k=0$ imply that the crossover value must be $s^*$. In this case, if we take the large deviation limit with the scaling $t$, as  in the previous sections, we find $I(s)=0$ for $s\geq s^*$, as shown in Fig.~\ref{figure1}(b). This result is also consistent with the divergence of $\lambda_{\Delta s_{\tot}}(k)$ for $k>0$, though we must stress again that a zero rate function is in general only an artifact of taking the large deviation limit with the ``wrong'' time scaling \cite{touchette2009} -- it is a signal that a ``fat'' LDP governs the large fluctuations of the entropy production, which for our model has the form shown in~(\ref{eqsubldp1}).

To find the LDP scaling associated with the \emph{negative}
fluctuations of the entropy production, we can develop a similar
argument by considering large, negative displacements brought about by
negative velocities over short-lived collisions. However, in this case
it is much simpler to combine the FR of Eq.~(\ref{eqpr1}) and the LDP (\ref{eqsubldp1}) to
obtain 
\be P(\Delta s_{\tot}/t=-s) \sim e^{-ts-\kappa
  \sqrt{t}\sqrt{s}}\sim e^{-ts},\qquad s\geq s^*.  
\ee 
Therefore, similarly to its
center, the pdf $P(\Delta s_{\tot}/t=s)$ satisfies
an LDP for $s\leq -s^{*}$ whose dominant exponential scale is $t$ and whose rate function at
this scale is $I(s)=-s$. This result is illustrated in
Fig.~\ref{figure1}(b) and is consistent with the
divergence of $\lambda_{\Delta s_{\tot}}(k)$ for $k<-1$. 

With this part, we now have the full entropy production pdf. To
summarize, we have found two LDPs for $P(\Delta s_{\tot}/t=s)$: one at speed $t$, which
correctly characterizes the large deviations of the entropy production for $s\leq s^*$, but not for 
$s>s^*$ since $I(s)=0$ in that region; and a second LDP at speed $\sqrt{t}$,
which refines the first one in the region $s>s^*$.

\section{Conclusions}

We have discussed in this paper the interplay that exists between the
fluctuation relation (FR) and the large deviation principle (LDP) for
a collisional Maxwell-Lorentz gas. We have used this model to
demonstrate that the FR is more fundamental in a sense than the
LDP. Focusing on the entropy production of this model, we have indeed
shown that, although this quantity satisfies an FR, the pdf of this
quantity involves two different large deviation scalings with time,
which implies that an FR does not necessarily arise, as often thought,
from a uniform LDP decaying everywhere exponentially with time. The
generality of the FR in this case enabled us to obtain corrections to
the exponential LDP, thereby showing that the FR can be used in a
constructive way to obtain large deviations when other approaches,
based for example on generating functions and the G\"artner-Ellis
Theorem, fail.

Although the model studied is greatly simplified compared to real
gases, it serves as a valuable benchmark to study granular
systems~\cite{ernst,baldassa} and conduction problems~\cite{beijeren},
among other physical phenomena.  Moreover, the fact that the
non-uniform LDP that we find is related to long ballistic
accelerations of the gas' particles shows that our results should
apply to more general collisional models, having for example a
dependence of the mean free time on the relative velocity between
particles. Recent studies of the hard-sphere case show that the
distribution of times between collisions conserves an exponential tail
at large times \cite{VWT08}, so that the mechanism discussed here should also be
relevant for this case. 

In closing the paper, we should mention that, since autocorrelation functions
are observed to decay exponentially for this model~\cite{gradenigo2012a}, we do not expect any
violations of the Green-Kubo relations. We should also note that our results
do not apply to \emph{thermostatted} Lorentz gas~\cite{DK05}, for which
a thermostat acts in between collisions, leading to additional contributions to the entropy 
production.

\appendix

\section{Calculation of the SCGFs}

\subsection{Work}
\label{appscgf}

We can expand from Eq.~(\ref{eq:work}) the expression of the SCGF of the work $W(t)$ in the following way:
\begin{eqnarray}
\lambda_W(k) &=& \lim_{t \to \infty} \frac{1}{t} \ln \left(
\sum_{n=0}^\infty p_t(n) \left\langle e^{\frac{k m
    \mE}{\theta} \sum_{i=1}^{n-1}x_i}\right\rangle\right)
\nonumber \\ 
&=&\lim_{t \to \infty} \frac{1}{t} \ln \left[
  e^{-\frac{t}{\tau_c}}\sum_{n=0}^\infty\frac{1}{n!}\left(\frac{t}{\tau_c}\right)^n
  \left\langle e^{\frac{k m \mE}{\theta} x}
  \right\rangle^{n-1} \right] \nonumber \\ 
&=& \lim_{t \to \infty} \frac{1}{t}
\ln \left[ \frac{e^{-\frac{t}{\tau_c}}}{G_x(k)}\sum_{n=0}^\infty
  \frac{1}{n!}\left(\frac{t G_x(k)}{\tau_c}\right)^n\right] \nonumber \\ 
&=& \frac{G_x(k)-1}{\tau_c}.
\label{eq:SMnew2}
\end{eqnarray}
where $G_x(k)$ is the generating function of the displacements $x_i$
defined after Eq.~(\ref{eq:SMnew}). In the second line, we have used
the statistical independency of the $n-1$ displacements $x_i$ and the
fact that the number $n$ of collisions in a time $t$ is distributed
according to the Poisson statistics:
\begin{equation}
p_t(n)=\frac{(t/\tau_c)^n}{n!} e^{-t/\tau_c}.
\end{equation}
The displacements are themselves random variables taken
from the following pdf:
\begin{equation}
P_x(x) = \sqrt{\frac{q}{\pi}} \int_{-\infty}^{\infty} dv\, e^{-q
  v^2} \int_{0}^{\infty} \frac{d\tau}{\tau_c}~e^{-\tau/\tau_c}
\delta[x-(v\tau + \tau^2\mE/2)],
\label{eq:pdx}
\end{equation}
so that, with $m/\theta=2 q$, we finally get 
\begin{eqnarray}
\hspace{-2cm}  G_x(k) =
\int_{-\infty}^\infty dx~e^{2kq\mE x} P_x(x) = \int_0^\infty \frac{d\tau}{\tau_c} \exp\left[-\frac{\tau}{\tau_c}
  + q \mE^2\tau^2 k\left(1+k\right)\right].
\label{eq:deltax}
\end{eqnarray}

\subsection{Boundary term}
\label{appscgf2}

To find the SCGF of the boundary term, we write it as the sum $B=b_1+b_n$ of the two terms defined in Eqs.~(\ref{bn}), which are asymptotically
independent, since they involve velocities that are
separated by a large number of independent random collisions. As a result, we can write
\be
\langle e^{kB}\rangle=\langle e^{k b_1}\rangle\langle e^{k b_n}\rangle
\label{eqsplit}
\ee
in the limit $t\ra\infty$. Of course, the velocity $v(0)$ at the initial time is correlated with 
the velocity  $v(t_1^-)$ before the first collision,
while the velocity $v(t_n^+)$ after the last collision
is correlated with the velocity $v(t)$ at the final time.  
In fact, we have
\begin{eqnarray}
v(t_1^-) = v(0) + \mE \tau \label{b1}\nonumber\\ 
v(t) = v(t_n^+) + \mE  \sigma, \label{b2}  
\end{eqnarray}
where $\tau$ and $\sigma$ are random variables distributed according
to the exponential pdf $P_\tau(\tau)$.
From this, it is easy to see that the probability distribution of the intervals
between an \emph{arbitrary} time, e.g., $t=0$ and the following
collision $t_1^-$, is the same as that between successive
collisions~\cite{haus}.  Using (\ref{b2}) above, we must then 
have
\begin{eqnarray} 
b_1 = \frac{m}{2\theta} (v+\mE\tau)^2 + \ln P_{\stat}(v) \nonumber \\
b_n = - \frac{m}{2\theta} u^2 - \ln P_{\stat}(-(u+\mE\sigma)), 
\end{eqnarray}
where $v$ has the stationary pdf (\ref{eq:stationary})
and $u$ has the Gaussian pdf (\ref{eq:gaussian}). From these pdfs, we thus write
\begin{eqnarray} 
\hspace{-2.0cm} \langle e^{k b_1}\rangle &=&\int_{-\infty}^{\infty} dv \int_{0}^{\infty} \frac{d\tau}{\tau_c}~e^{-\tau/\tau_c}~P_{\stat}^{k+1}(v) e^{k q (v+\mE t)^2} \label{int1} \\
\hspace{-2.0cm} \langle e^{k b_n}\rangle &=&\int_{-\infty}^{\infty} du \int_{0}^{\infty} \frac{d\tau}{\tau_c}~e^{-\tau/\tau_c}~P_{\stat}^{-k}(-u-\mE\tau)P_{\scatt}(u)e^{-k q u^2}, \label{int2}
\end{eqnarray}
where $q=m/2\theta=M/2T$.

At this point, we can use the asymptotic behavior of $P_\stat(v)$, given by
\be
P_\stat(v)\sim\left\{  
\begin{array}{ll}
e^{-bv} & \textrm{for $v\to+\infty$} \\
e^{-qv^2} & \textrm{for $v\to-\infty$,} 
\end{array} \right.
\ee
to find the regions of convergence of the integrals~(\ref{int1})
and~(\ref{int2}).  For the first integral, we find $k\in (-\infty,0]$, while for the second one we have $k\in (-1,0]$. These domains of convergence  determine the domain of convergence of $\lambda_{\Delta s_{\tot}}(k)$ via Eqs.~(\ref{eqsplit}) and (\ref{eq:DStot-expr}).

\section*{Acknowledgments}

H.T. thanks Rosemary J.\ Harris for comments on the manuscript. 
G.G., A.S.\ and A.P.\ acknowledge support from the Italian MIUR FIRB-IDEAS program (grant no.\ RBID08Z9JE). A.P.\ also acknowledges support from the Italian MIUR PRIN program (grant no.\ 2009PYYZM5). 

\section*{References}

\bibliography{ldpfr}
\bibliographystyle{unsrt}

\end{document}